\documentclass[twocolumn,aps,prl,superscriptaddress]{revtex4}
\usepackage{amsfonts}
\usepackage{amsmath}
\usepackage{txfonts}
\usepackage{amssymb}
\usepackage{bbold}
\usepackage{amsbsy} 
\usepackage{epsfig}
\usepackage{cancel}
\usepackage{float}
\usepackage{romannum}
\usepackage{color}

\def\bea{\begin{eqnarray}} \def\eea{\end{eqnarray}}

\def\bsigma{{\boldsymbol \sigma}}

\def\bq{{\bf q}}

\def\bk{{\bf k}}
\def\br{{\bf r}}
\def\bd{{\bf d}}

\def\bg{{\bf g}}
\def\bp{{\bf p}}

\def\bS{{\bf S}}

\def\bn{\hat{\bn}}

\begin{document}

\title{Double helix nodal line superconductor}

\author{Xiao-Qi Sun}
\affiliation{Department of Physics, McCullough Building, Stanford University, Stanford, California 94305-4045, USA}
\author{Biao Lian}
\affiliation{Department of Physics, McCullough Building, Stanford University, Stanford, California 94305-4045, USA}
\author{Shou-Cheng Zhang}
\affiliation{Department of Physics, McCullough Building, Stanford University, Stanford, California 94305-4045, USA}
\begin{abstract}
Time-reversal invariant superconductors in three dimensions may contain nodal lines in the Brillouin zone, which behave as Wilson loops of 3D momentum-space Chern-Simons theory of the Berry connection. Here we study the conditions of realizing linked nodal lines (Wilson loops), which yield a topological contribution to the thermal magnetoelectric coefficient that is given by the Chern-Simons action. We find the essential conditions are the existence of torus or higher genus Fermi surfaces and spiral spin textures. We construct such a model with two torus Fermi surfaces, where a generic spin-dependent interaction leads to double-helix-like linked nodal lines as the superconductivity is developed.
\end{abstract}
\maketitle

The nodal-line superconductor is an intriguing class of three dimensional (3D) unconventional superconductor that respects the time-reversal symmetry. Unlike usual superconductors which are fully gapped, a nodal-line superconductor contains closed gapless nodal lines in the Brillouin zone (BZ), which are protected by time-reversal symmetry. In nature, such superconductors widely occur in cuprates \cite{hu1994,wollman1993,wollman1995,kirtley1995}, iron-based superconductors \cite{okazaki2012,zhang2012} and noncentrosymmetric superconductors \cite{bauer2004,yip2014}. As 1D loops, the nodal lines in the 3D BZ can form nontrivial links\cite{lian2017,chen2017,yan2017,ezawa2017}, where the linking numbers are topologically invariant and give a classification of nodal-line superconductors \cite{lian2017}. Physically, change of linking numbers of nodal lines has been shown to yield a topological shift in the coefficient of thermal magnetoelectric effect, which is also known as the theta angle \cite{lian2017,wang2011,ryu2012}. This is because the theta angle is theoretically given by the action of the Chern-Simons (CS) theory of Berry connection in the 3D BZ \cite{wang2011}, while the nodal lines behave exactly as Wilson loops in the theory \cite{lian2017}, whose linking numbers contributes topologically to the CS action as first shown by Witten \cite{polyakov1988,witten1989}. On the surface of the superconductor, linked nodal lines lead to topologically protected kissing Majorana flat bands  which are bounded by the projected nodal lines \cite{schnyder2011,schnyder2012,schnyder2015,sato2011,tanaka2011}.
However, superconductors with linked nodal lines have not been discovered so far. In this Letter, we investigate the physical conditions for realizing linked nodal-line superconductors, and we show the key ingredients for linked nodal lines to occur are torus or higher genus Fermi surfaces and certain spiral spin textures. In particular, we construct a double helix nodal-line superconductor lattice model, which contains two pairs of linked nodal lines resembling the DNA double helix structure \cite{watson1953,watson2012} under the standard BCS mean field theory. Our theory gived a useful guidance on the search for linked nodal-line superconductors in nature.

Though nodal lines could occur in centrosymmetric TRI materials such as cuprates, they appear more commonly in noncentrosymmetric TRI superconductors, where the electron bands are generically nondegenerate due to noncentrosymmetric spin-orbital couplings (SOCs) \cite{anderson1984,yip2014}. Here we restrict ourselves to the noncentrosymmetric TRI superconductors, which allows us to consider a larger variety of Hamiltonians.
The stability of nodal lines is protected by the time-reversal symmetry.
Given an electron band of dispersion $\epsilon(\mathbf{k})$ where $\mathbf{k}$ is the lattice momentum, the time-reversal symmetry restricts its pairing amplitude $\Delta(\mathbf{k})$ to be real in a TRI basis \cite{qi2009}. Therefore, the system is gapless when the two conditions $\epsilon(\mathbf{k})=\Delta(\mathbf{k})=0$ are satisfied, which gives rise to one-dimensional nodal lines in the BZ.
Note that the nodal lines are restricted on the Fermi surface of the band given by $\epsilon(\mathbf{k})=0$, so the Fermi surface topology will limit the possible configurations of nodal lines.

\begin{figure}
\includegraphics[width=0.5\textwidth]{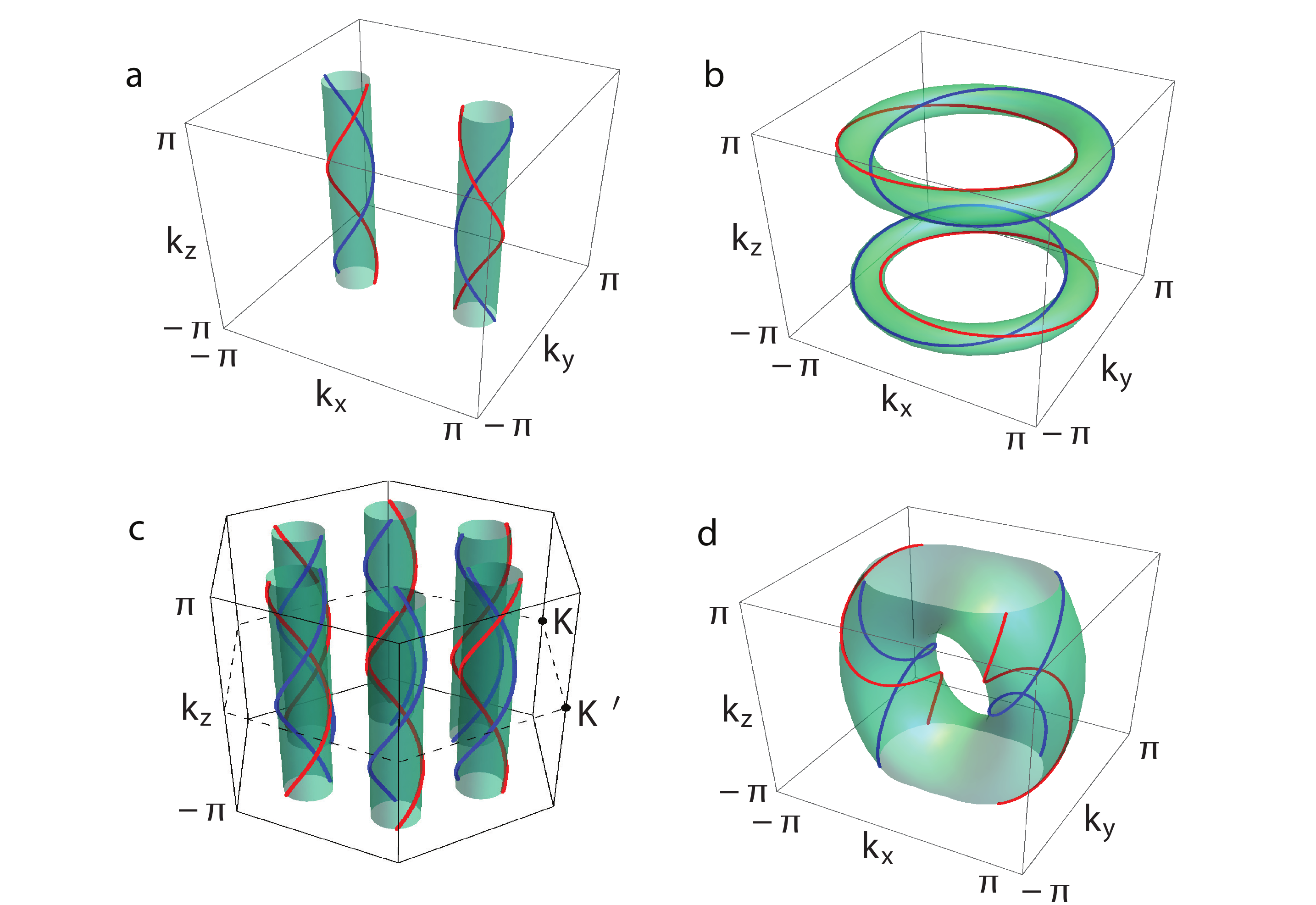}
\caption{Examples of Fermi surfaces compatible with TRI linked nodal lines resembling the double helix. (a)-(c) shows how such linked nodal lines may occur on torus Fermi surfaces.
(d) shows a possible configuration of linked nodal lines on a genus two Fermi surface.}
\label{fermi-surfaces}
\end{figure}

It is easy to see that a spherical Fermi surface does not support any kind of linked nodal lines.
While if the Fermi surface is a torus, one is able to draw two nodal lines linked with each other along the toroidal direction of the torus, which resembles a closed DNA double helix. An integer Gauss linking number\cite{gauss1867,polyakov1988,witten1989} $n_L$ can be defined for such two nodal lines, which is equal to the total number of coils of the double helix (up to a sign).
However, one must note the linking number $n_L$ reverses sign under time-reversal transformation, so a double helix cannot be TRI by itself. Instead, the superconductor must contain minimally two double helices which are time reversal partners of each other, and this requires at least two torus Fermi surfaces forming a time reversal pair. Figure \ref{fermi-surfaces}(a) shows such a configuration of two double helices with linking number $n_L=\pm1$, respectively, where the two torus Fermi surfaces are cylinders periodic in the $k_z$ direction. For Fermi surfaces with a higher genus (number of holes) $N_{g}\ge2$, it is possible to draw two such double helices on a single TRI Fermi surface, and Fig.\ref{fermi-surfaces}(d) shows such an example. In nature, Fermi surfaces with a high genus are quite common in metals.

Given a suitable topology of Fermi surfaces, we still need a nontrivial electron-electron interaction to achieve linked nodal lines in the superconductor. Here we take the example of Fig.\ref{fermi-surfaces}(a) where two cylindrical torus Fermi surfaces are centered at $(k_x,k_y)=(\pm Q_x,0)$, and investigate how the double-helix linked nodal lines of linking number $n_L=\pm1$ can be realized. The Cooper channel interaction of the superconductor is generically given by $H_I=\sum_{\bk,\bk'}(V_{\bk\bk'}/2N)c^{\dagger}_{\bk}c^{\dagger}_{-\bk}c_{-\bk'}c_{\bk'}$, where $c_{\bk}$ and $c_{\bk}^\dag$ are the electron annihilation and creation operators at momentum $\bk$ near the Fermi surfaces, $V_{\bk\bk'}$ are the interaction coefficients, and $N$ is the number of unit cells.
In the presence of time-reversal symmetry, the pairing amplitude $\Delta_{\bk}=\langle \sum_{\bk'}V_{\bk\bk'} c_{-\bk'}c_{\bk'}\rangle/N$ is real. The mean-field free energy of the superconductor is then $F(\Delta_{\bk})= \sum_{\bk,\bk'}(\Delta_{\bk}/2E_\bk)V_{\bk\bk'}(\Delta_{\bk'}/2E_{\bk'})/2N$, where $E_\bk=(\epsilon_\bk^2+\Delta_\bk^2)^{1/2}$ is the Bogoliubov-de Gennes quasiparticle spectrum. Therefore, $\Delta_\bk$ and $\Delta_{\bk'}$ tend to have the same (opposite) sign for $V_{\bk\bk'}<0$ ($V_{\bk\bk'}>0$). By definition of nodal lines, $\Delta_\bk$ on a Fermi surface will take opposite signs on the two sides of a nodal line.
For the double helix nodal lines shown in Fig.\ref{fermi-surfaces}(a), one finds $\Delta_\bk\Delta_{\bk'}\le0$ when $\bk-\bk'=(0,0,\pi)$, or when $\bk$ and $\bk'$ lie in the same plane of constant $k_z$ and opposite to each other on the same cylindrical Fermi surface. 
This motivates us to write down the following interaction for $\bk$ and $\bk'$ on the same Fermi surface:
\begin{equation}
V_{\bk\bk'}=-\left(\eta_{1}\widetilde{k_{x}}\widetilde{k_{x}'}+\eta_{2}k_{y}k_{y}'\right) \cos(k_{z}-k_{z}')\ ,
\label{prototype}
\end{equation}
where we have defined $\widetilde{k_x}=k_x\mp Q_x$ for momentum $\bk$ near $(k_{x},k_{y})=(\pm Q_{x},0)$, respectively. The interaction at the other momentums can be obtained via the relation $V_{\bk\bk'}=-V_{\bk,-\bk'}=-V_{-\bk,\bk'}$.
We expect the double helix nodal lines with linking number $\pm1$ to be favored when such a term dominates in the electron-electron interaction near Fermi surfaces.

Such a $\bk$-dependent interaction can be realized from a spin-dependent Heisenberg interaction together with a spiral Fermi-surface spin texture due to SOC. To illustrate this idea, we can start from two spin-degenerate cylindrical Fermi surfaces at $(k_x,k_y)=(\pm Q_x,0)$, and add a SOC as follows:
\begin{equation}
H_{\text{soc}}(\bk)=\bg(\bk)\cdot \bsigma=\lambda(k_{y} \sigma_{x}-\widetilde{k_{x}}\sigma_{y})+\lambda_{z}(\sin k_{z}\sigma_{x}\pm \cos k_{z}\sigma_{y}),
\label{SOC}
\end{equation}
where $\sigma_{x,y,z}$ are the Pauli matrices for spins, the $\pm$ signs correspond to $\bk$ near $(k_{x},k_{y})=(\pm Q_{x},0)$, respectively, and we keep only the leading order expansion in $\widetilde{k_x}$ and $k_y$. Because of the SOC, the Fermi surfaces split into two inner cylinders and two outer ones centered at $(k_{x},k_{y})=(\pm Q_{x},0)$.
In particular, the $\lambda_z$ term leads to a spiral spin texture on each Fermi surface that rotates along the $k_z$ direction. For a generic spin-dependent interaction $H_I=V(\br_1-\br_2)\rho(\br_1)\rho(\br_2)+J(\br_1-\br_2)\bS(\br_1)\cdot\bS(\br_2)$ where $\rho(\br)$ and $\bS(\br)$ are the local electron density and spin operators.
The projection onto the spin texture will naturally yield an interaction as shown in Eq.(\ref{prototype}), as we will show later.

\begin{figure}
\centering
\includegraphics[width=0.5\textwidth]{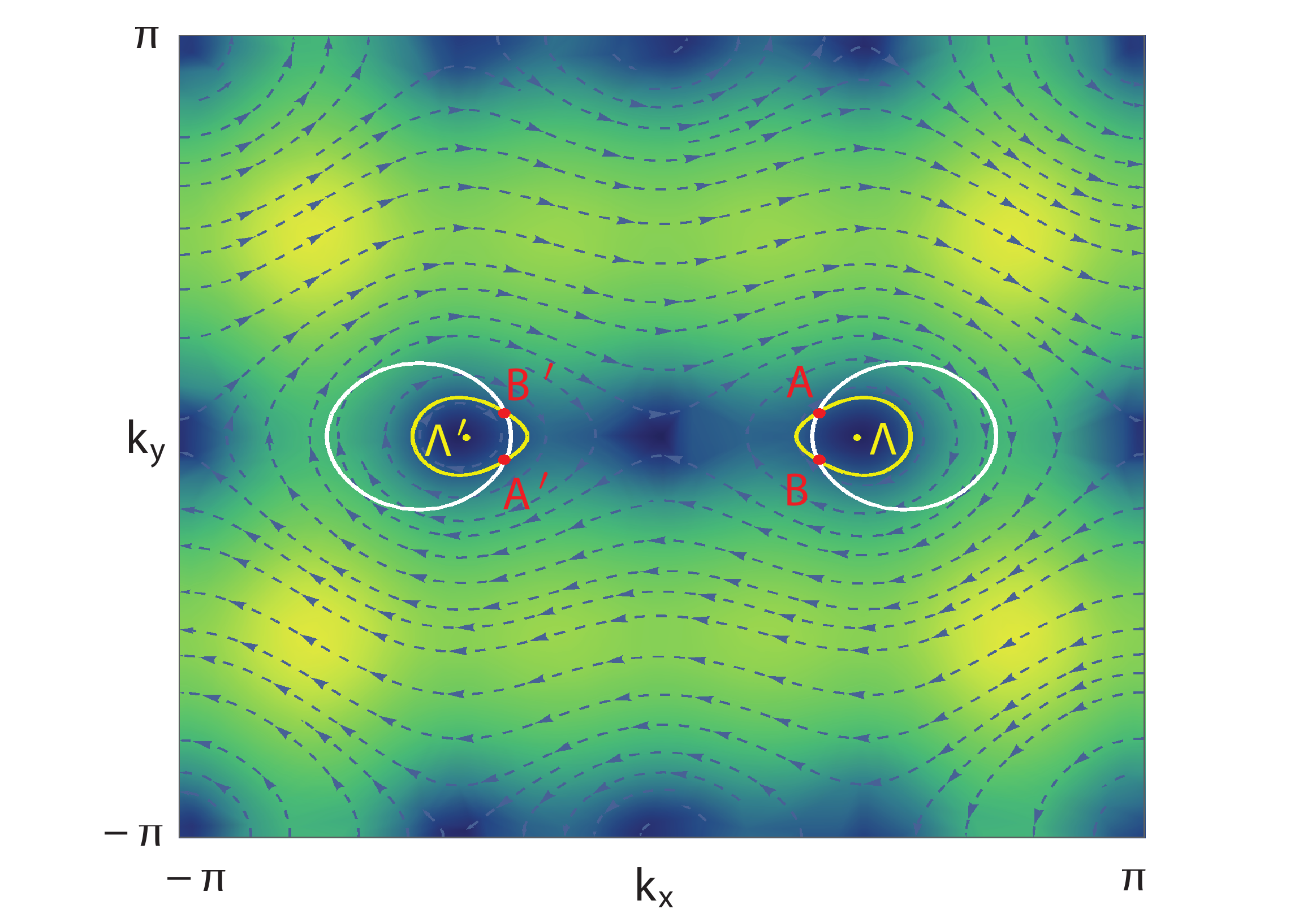}
\caption{The arrows show the spin textures of the lower electron band in the $k_z=0$ plane for $\lambda=-0.02 t$, $\lambda_z=0.005 t$, and the brightness is proportional to the SOC strength $|\bg(\bk)|$ (brighter color for larger $|\bg(\bk)|$). In particular, $|\bg(\bk)|=0$ at $\Lambda$ and $\Lambda'$ points. The nodal lines occur at $A$, $B$ and $A'$, $B'$ points which are intersections between the Fermi surfaces as given by the white circles and the constant $|g(\bk)|=\psi/\xi$ contours (yellow circles).}
\label{spin-texture}
\end{figure}

We also want to mention a different way to understand the linked nodal lines induced by SOC as follows. Generically, the TRI pairing amplitude of the two bands split by SOC is a matrix in the natural electron spin $s_z=\pm1/2$ basis as follows \cite{sigrist1991}:
\begin{equation}
\Delta_{ss'}(\bk)= \left\{\left[\psi(\bk)+\bd(\bk)\cdot \bsigma\right]i\sigma_{y}\right\}_{ss'}\ ,
\end{equation}
where $\psi(\bk)=\psi(-\bk)$ gives a singlet pairing while $\bd(\bk)=-\bd(-\bk)$ corresponds to a triplet pairing, both of which are real functions of $\bk$. For a system with a SOC given by $H_{\text{soc}}=\bg(\bk)\cdot \bsigma$, it is shown \cite{frigeri2004} that the most favorable pairing amplitude satisfies $\bd(\bk)=\xi(\bk)\bg(\bk)$, where $\xi(\bk)$ is a scalar function. This yields a pairing amplitude  $\Delta(\mathbf{k})=\psi(\bk)\mp\xi(\bk)|\bg(\bk)|$ on the two outer ($-$ sign) and inner ($+$ sign) cylindrical Fermi surfaces. In the simplest case when both $\psi(\bk)=\psi$ and $\xi(\bk)=\xi$ are constants, nodal lines will occur at the intersections between the outer cylindrical Fermi surfaces and the constant $|\bg(\bk)|$ surfaces defined by $\Delta(\bk)=\psi-\xi|\bg(\bk)|=0$. Figure \ref{spin-texture}(a) shows the intersection nodes $A$, $B$ and $A'$, $B'$ in the $k_z=0$ plane, where the white circles are the outer cylindrical Fermi surfaces, and the yellow loops are constant $|\bg(\bk)|=\psi/\xi>0$ surfaces, which are centered at $\Lambda$ and $\Lambda'$ where $|\bg(\bk)|=0$. The light (dark) color represents the regions where $|\bg(\bk)|$ is large (small), and the arrows show the spin textures in the BZ. As $k_z$ increases from $0$ to $2\pi$, the $\Lambda$ point moves along a spiral trajectory $\left(Q_{x}+\frac{\lambda_z}{\lambda}\cos k_z,-\frac{\lambda_z}{\lambda} \sin k_z, k_z\right)$, and similarly for $\Lambda'$, so the gapless nodes $A$, $B$ (and $A'$, $B'$) will undergo a $2\pi$ rotation with respect to $k_z$, yielding a double helix nodal-line structure. We note that in order to have intersection nodes, the value of $\psi/\xi$ must be within a certain range, which is determined by interactions. Similarly, nodal lines may also occur on the inner Fermi surfaces at intersections where $|\bg(\bk)|=-\psi/\xi$ (if $\psi/\xi<0$), which also form double helices.

Now we rewrite the above model in a 3D lattice, and demonstrate the above interaction and SOC indeed give rise to double-helix linked nodal lines. We assume the following single-electron Hamiltonian with SOC:
\begin{equation}
\begin{split}
H_{0}(\bk)&=t'\cos 2k_x-t \cos k_y +\lambda_{1}\sigma_x \sin k_y+\lambda_2\sigma_y\sin 2k_x\\
&+\lambda_{3}\sigma_{x}\sin k_z+\lambda_4 \sigma_y \sin k_x \cos k_z-\mu\ ,
\end{split}
\label{lattice}
\end{equation}
where $t$ and $t'$ are positive, and $\lambda_i$ ($1\le i\le 4$) are SOC parameters which are TRI but inversion asymmetric. When the SOC is zero, the electron kinetic energy has its minima at $(k_x,k_y)=(\pm \pi/2,0)$. For chemical potential $\mu$ satisfying $-t'-t<\mu<-|t-t'|$, the Fermi surfaces are two spin-degenerate cylindrical tori centered at the minima. When the SOC is turned on, the Fermi surfaces split into two inner tori and two outer tori, and acquire spin textures. For convenience, we fix $\lambda_1=2\lambda_2=\lambda$ and $\lambda_3=\lambda_4=\lambda_z$. The SOC reduces to the form of Eq.(\ref{SOC}) in the vicinity of $(k_x,k_y)=(\pm \pi/2,0)$. In principle, the Hamiltonian could contain other hopping or SOC terms, but we will keep them zero since they do not qualitatively change the physics \cite{suppl}. In the following discussion, we will fix the single-particle parameters at $t'=0.25 t$, $\lambda=-0.2t$, $\lambda_z=0.4t$. For simplicity, we will also fix the chemical potential at $\mu=-1.6t$ slightly above the band minimum, so that the two inner torus Fermi surfaces shrink to zero \cite{suppl}, with only the two outer torus Fermi surfaces left.

We then add an electron-electron interaction to the system as follows\cite{sigrist2005}:
\begin{equation}
H_{\text{int}}=\frac{1}{N}\sum_{\bq} V(\bq)\rho_{\bq} \rho_{-\bq}+\frac{1}{N}\sum_{\bq}J(\bq)\bS_{\bq}\cdot \bS_{-\bq}\ ,
\end{equation}
where $\rho_\bq=\sum_{\bk s}c_{\bk+\bq,s}^{\dagger} c_{\bk,s}$ and $\bS_{\bq}=\frac{1}{2}\sum_{\bk ss'} c_{\bk+\bq,s}^{\dagger} \bsigma_{ss'}c_{\bk, s'}$ are the electron density and spin operators at momentum $\bq$. Such an interaction may generically arise from the phonon exchange and the electron itinerant magnetism. Explicitly, we assume the interaction potentials take the form $V(\bq)=V_0+V_x \cos 2 k_x+ V_y \cos k_y$ and $J(\bq)=J_x \cos 2k_x+J_y \cos k_y$, where $V_0, V_x, V_y, J_x$ and $J_y$ are constants. With the spin textures known, one can readily project the interaction onto the two cylindrical Fermi surfaces. Using a $\bk\cdot \bp$ expansion, one can show the Cooper channel interaction potential $V_{\bk\bk'}$ for $\bk$ and $\bk'$ on the same Fermi surface is given by \cite{suppl}
\begin{equation}
\begin{split}
V_{\bk\bk'}&=\eta_{0}-\left(\eta_{1} \widetilde{k_x}\widetilde{k_x'}+\eta_{2} k_y k_y'\right)\cos\left[\phi(\bk)-\phi(\bk')\right]\\
&-\eta_3(\widetilde{k_x}^2+\widetilde{k_x'}^2)-\eta_4(k_y^2+k_y'^2)\ .
\end{split}
\label{interaction}
\end{equation}
where we have kept terms up to the second order in $\widetilde{k_x}$ and $k_y$. Here $\phi(\bk)$ is the angle of the spin direction at $\bk$ in the $k_x$-$k_y$ plane, $\eta_1=-(4V_x+J_x)$, $\eta_2=-(V_y+J_y/4)$, $\eta_3=2V_x-3J_x/2$, $\eta_4=V_y/2-3J_y/8$, $\eta_0=V_0+\eta_3/2+2\eta_4$, and we have $\widetilde{k_x}=k_x\mp \pi/2$ for momentum near $(k_x,k_y)=(\pm\pi/2,0)$, respectively. Compared with Eq.(\ref{prototype}), the above interaction contains a few other less important terms and has the spin angle $\phi(\bk)$ in place of $k_z$. However, since the spin angle $\phi(\bk)$ increases(decreases) by $2\pi$ as $k_z$ goes from $-\pi$ to $\pi$ on the two Fermi surfaces, the above interaction still satisfies the condition for double helix nodal lines. Based on this interaction near Fermi surfaces, we solve numerically the linearized gap equation that determines the mean-field superconductivity critical temperature $T_c$ \cite{sigrist2005}:
\begin{equation}
-\chi \Psi(\bk)=\frac{1}{2N\epsilon_c}\sum_{\bk'} V_{\bk \bk'} \Psi(\bk')\ ,
\label{TcEq}
\end{equation}
where $\epsilon_c$ is the interaction cutoff energy (e.g. Debye energy), $\mathbf{k}$ is summed within the energy shell $[-\epsilon_c,\epsilon_c]$ near the Fermi surface, while $\Psi(\bk)\propto\Delta(\bk)$ is an eigenvector of $V_{\bk \bk'}$, and $\lambda$ is the eigenvalue. For $\chi>0$, the critical temperature is given by $k_B T_c=1.14\epsilon_{c}e^{-1/\chi}$. For $\chi\le 0$, the eigenvector $\Psi(\bk)$ is not a pairing instability. The most favorable pairing function $\Delta(\bk)$ is proportional to the eigenvector $\Psi(\bk)$ with the largest positive eigenvalue $\chi$.

\begin{figure}
\centering
\includegraphics[width=0.49\textwidth]{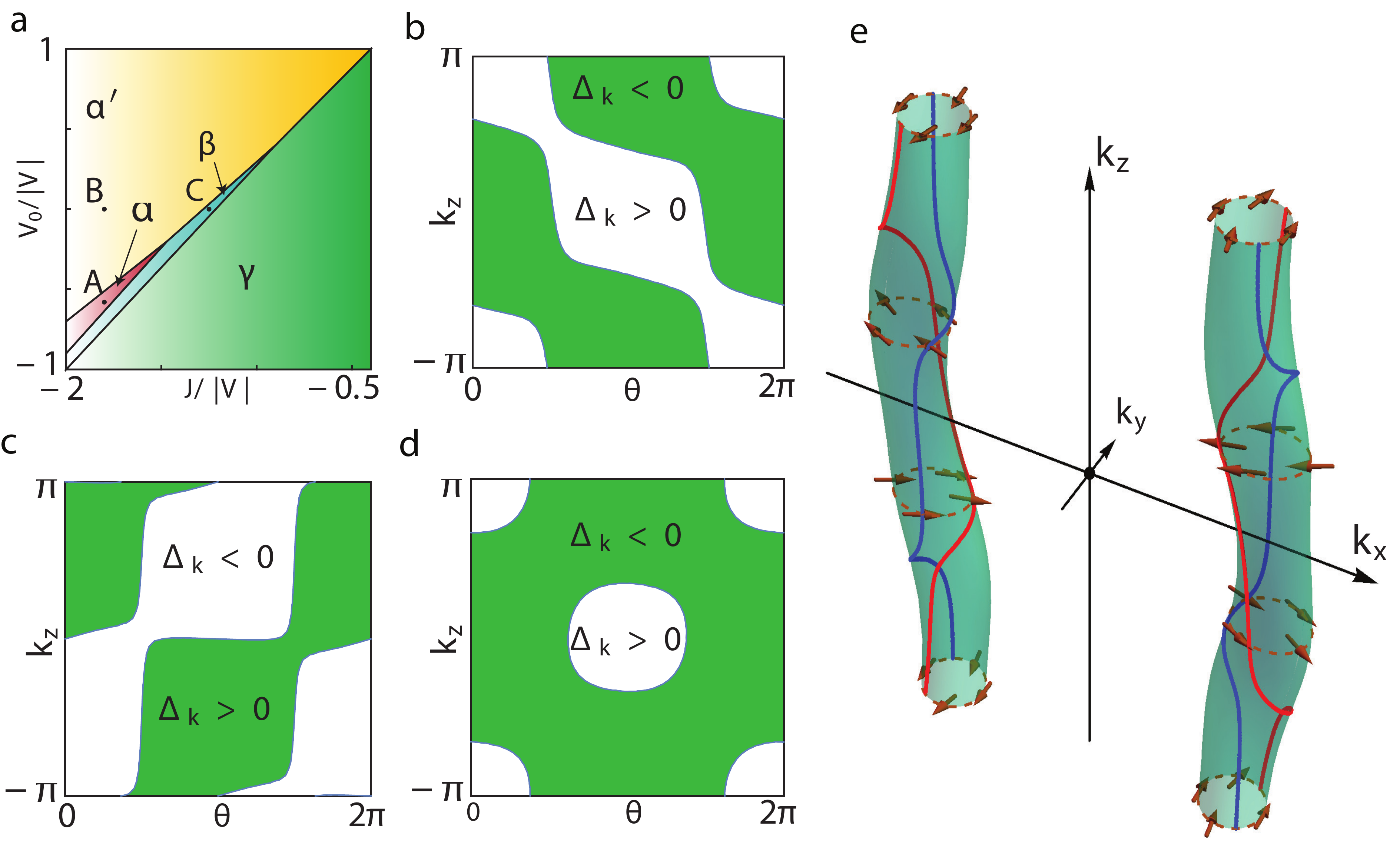}
\caption{(a) Phase diagram for fixed $V<0$, where the horizontal and vertical axes are $J$ and $V_0$, respectively. Both $\alpha$ and $\alpha'$ phases are double-helix linked nodal line phases, except that their helicities are opposite. The $\beta$ phase contains two unlinked nodal lines on each Fermi surface. The $\gamma$ phase is the conventional fully gapped phase. Panels (b)-(d) show the nodal-line configuration on the Fermi surface at $k_x=\pi/2$ for points $A$, $B$ and $C$ in the phase diagram, respectively, where $\theta=\arg(\widetilde{k_x}+ik_y)$ is the poloidal angle of the torus Fermi surface. (e)\ The 3D plot of the spin texture (arrows) and nodal lines (blue and red lines) on the Fermi surfaces at point $A$ of the phase diagram.}
\label{phase-diagram}
\end{figure}

Figure \ref{phase-diagram}(a) shows a phase diagram with respect to $J_x=J_y=J$ and $V_0$, where we have fixed $V_x=V_y=V<0$.
Different phases are distinguished by the topology of their nodal lines, and are labeled using different Greek letters. Both the $\alpha$ phase and the $\alpha'$ phase are double-helix nodal-line phases, while they have different double helix helicities: the $\alpha$ phase has a left-handed double helix at $k_x=\pi/2$, while the $\alpha'$ phase has a right-handed double helix at $k_x=\pi/2$. The typical nodal-line shapes of the $\alpha$ and $\alpha'$ phases are shown in Figs.\ref{phase-diagram}(b) and \ref{phase-diagram}(c), respectively, where $\theta=\arg(\widetilde{k_x}+ik_y)$ is the poloidal angle of the torus Fermi surface. Figure \ref{phase-diagram}(e) shows the double-helix nodal lines of the $\alpha$ phase in the 3D BZ.
We note that the physical picture in Fig.\ref{spin-texture} only predicts the double-helix helicity of phase $\alpha$, which is only valid for small $\lambda_{z}/\lambda$ \cite{suppl}. In the case we calculated in Fig.\ref{phase-diagram} where $\lambda_z/\lambda$ is large, both phases $\alpha$ and $\alpha'$ arise, which is because the projected interaction in Eq.(\ref{interaction}) in this case does not give a double-helix helicity preference.
The $\beta$ phase has two unlinked nodal loops as shown in Fig.\ref{phase-diagram}(d), while the $\gamma$ phase is fully gapped without any nodal lines.
The phase diagram shows that ferromagnetic spin interactions $J<0$ tend to favor double-helix nodal lines, which makes $\eta_1$ and $\eta_2$ positive as required in our argument. We note that ferromagnetic electron interactions are quite common for metals. Besides, $V_0$ cannot be too negative, otherwise the conventional fully gapped phase $\gamma$ will be more favorable.

\begin{figure}
\centering
\includegraphics[width=0.5\textwidth]{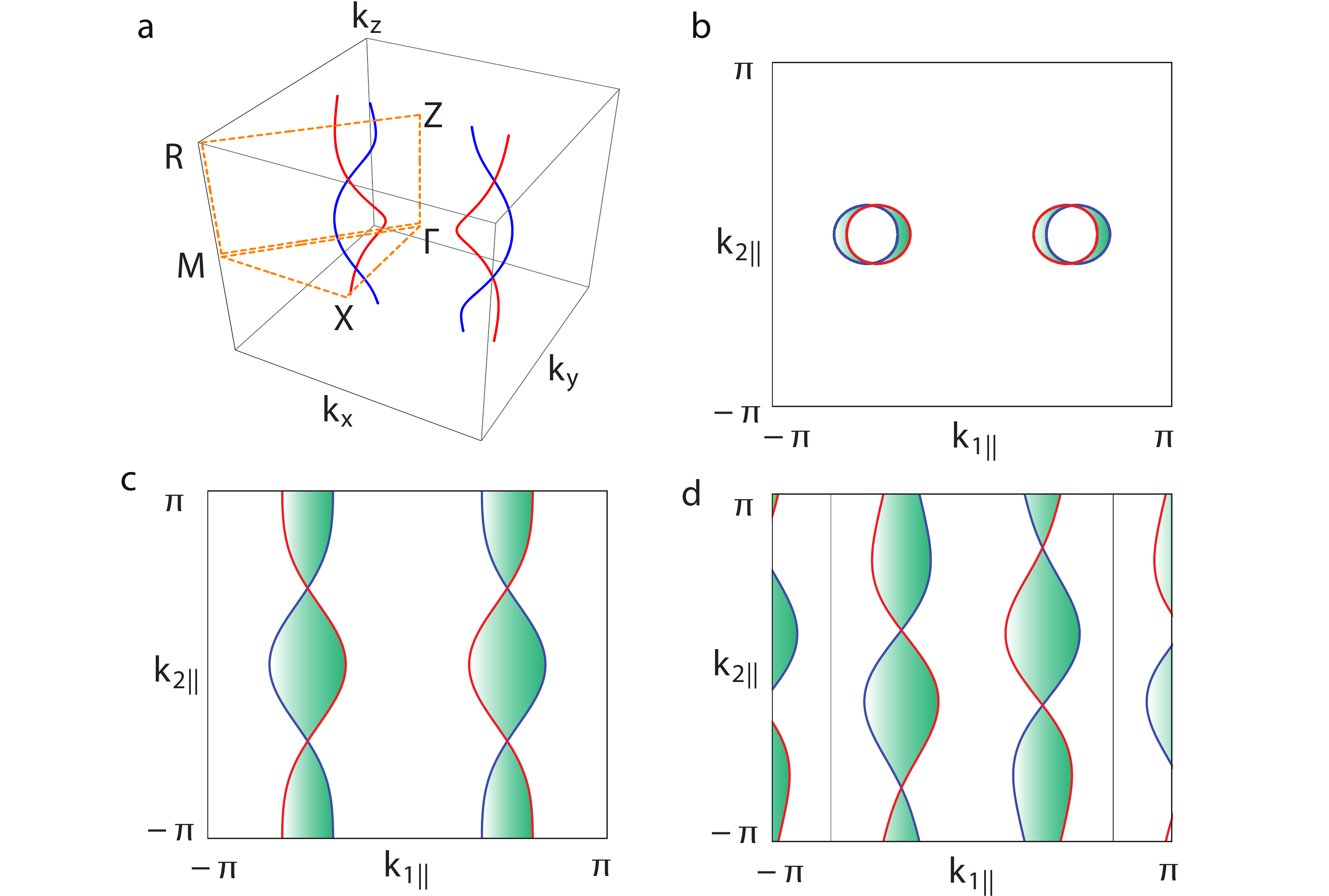}
\caption{(a) Sketch of a typical example of double-helix nodal lines in the BZ. (b)-(d) Illustration of the zero-energy surface Majorana flat bands (colored in green) on surface cuts along the $\Gamma$-$Z$, $\Gamma$-$X$, and $\Gamma$-$M$ directions, which are bounded by the projection of nodal lines. For linked nodal lines, the flat bands always kiss each other no matter which surface cut one takes. }
\label{surface-states}
\end{figure}

We have seen that the realization of linked nodal lines requires torus or higher genus Fermi surfaces, and strongly varying spin textures along the toroidal direction of the Fermi surfaces. These conditions put a limitation on the crystalline symmetry of candidate materials. In general, a lower crystalline symmetry is preferred. In particular, in the double helix nodal line configuration in Fig.\ref{fermi-surfaces}a, the only allowed point-group symmetry (up to a translation) is either a twofold rotation $C_{2x}$ about $x$ axis or a mirror reflection $M_x$ from $x$ to $-x$.
In other configurations as shown Figs.\ref{fermi-surfaces}(b)-1(d), a few other symmetry operations such as the threefold rotation $C_{3z}$ are allowed. For instance, the configuration of Fig.\ref{fermi-surfaces}(c) is allowed for materials with a hexagonal 3\textit{R} structure, which is common in transition metal dichalcogenides\cite{chhowalla2013}. We therefore suggest to search for (double-helix) linked nodal-line superconductors in layered metallic materials with strong SOCs and low crystalline symmetries, such as lithium-inserted metal oxide, transition metal dichalcogenides and transition metal halides, etc\cite{miro2014}, where torus Fermi surfaces and strongly varying spin textures are more likely to occur.

The detection of linked nodal lines is also essential in experiments. A few techniques such as corner junction \cite{wollman1993,van1995}, angular-resolved thermal transport measurement \cite{graf1996,matsuda2006} and angular-resolved photoemission spectroscopy (ARPES)\cite{zhang2012} have been used to reveal the nodal structure of superconductors, which will, however, be more complicated for linked nodal lines. Instead, we suggest a possibly easier way to reveal linked nodal lines through probing the surface states on the surface cuts of different directions using ARPES. As we mentioned at the beginning, nodal line superconductors exhibit topologically protected zero energy Majorana flat bands, which are bounded by the projection of nodal lines on the surface \cite{schnyder2011,schnyder2012,schnyder2015}. If two nodal lines are linked, their projections will necessarily cross each other on any surface cuts. Therefore, the corresponding surface Majorana flat bands always kiss each other as shown in Fig.\ref{surface-states}, no matter which surface cut one takes.


In conclusion, we have demonstrated the possibility of realizing linked nodal lines in noncentrosymmetric superconductors with torus or higher genus Fermi surfaces and strongly $\bk$-dependent spin textures, and studied an explicit lattice model that realizes a double-helix nodal-line superconductor. These results may serve as a preliminary guidance on searching for and detecting linked nodal-line superconductors in nature.
\begin{acknowledgements}
We are grateful to Peizhe Tang for helpful discussions. This work is supported by the NSF Grant No. DMR-1305677.
\end{acknowledgements}

\setcounter{figure}{0}    
\renewcommand{\figurename}{{\textbf{Supplementary Figure}}}
\renewcommand\thesubsection{\Alph{subsection}}
\appendix
\begin{widetext}
\section*{Supplemental Material}
In the main text, we have discussed double helix nodal line superconductor phase on a time-reversal pair of fermi surfaces shown in Fig.1a. A key ingredient is to have spiral spin texture in Fig.3(a)e. Such fermi surfaces and spin texture can be realized by lattice model Eq.(4) with proper spin-orbital coupling(SOC). We show the spin texture in Fig.2 and argue the existence of double helix nodal line superconductor from the relation of triplet pairing order parameter and SOC Hamiltonian. However, this physical picture is only valid at small $\lambda/\lambda_z$ when $|\bg(\bk)|=0$ points $\Lambda$ and $\Lambda'$ do exist near fermi surfaces. As a complement, we show the phase diagram of the small $\lambda/\lambda_z$ in subsection \ref{small-kz-dependent} to provide numeric evidence.  For large $\lambda/\lambda_z$ case, a proper way of understanding the phase of double helix nodal line superconductor in Fig.3(a) is via projected interaction on fermi surface. One form of projected interaction has been proposed as Eq.(1). In this supplemental material, we will support our arguments in the main text by presenting the calculation of fermi surface and projected interaction of lattice model Eq.(4-5) in subsection \ref{fermi-surface} and subsection \ref{projected-interaction}. We also justify two simplifications that we have made in the main text. The phase diagram Fig.3(a) is presented based on the gap function near critical temperature instead of zero temperature. We shall justify the method in subsection \ref{zero-temperature}. In the lattice model Eq.(4), we neglect other symmetry allowed hopping or SOC terms. We shall show in subsection \ref{spin-canting} that such terms as perturbation would generically not change the linking number of nodal lines.

\subsection{Fermi surface of single particle Hamiltonian}\label{fermi-surface}
In this supplemental subsection, we shall compute the equation of fermi surface for Hamiltonian Eq.(4). We rewrite the Hamiltonian as
\begin{equation}\tag{S1}
H(\bk)=t'\cos 2 k_x-t \cos k_y-\mu+g_x(\bk)\sigma_x+g_y(\bk)\sigma_y\ ,
\end{equation}
where we assume $t$ and $t'$ are positive and $g_x(\bk)=\lambda \sin k_y+\lambda_z \sin k_z$, $g_y(\bk)=\frac{1}{2} \lambda \sin 2 k_x+\lambda_z \sin k_x \cos k_z$  following the convention of main text. We diagonalize the Hamiltonian for dispersion relation:
\begin{equation}\tag{S2}
\epsilon_{\bk,\pm}=t'\cos 2 k_x-t \cos k_y-\mu \pm \sqrt{g_x(\bk)^2+g_y(\bk)^2}\ ,
\end{equation}
The label $\pm$ denotes upper or lower band electrons. The conditions of electron states of momentum $\bk$ being occupied and unoccupied are $\epsilon_{\bk,\pm}<0$ and $\epsilon_{\bk,\pm}>0$ respectively. Note that an inequality holds for all electron states in upper band:
\begin{equation}\tag{S3}
\epsilon_{\bk,+}\ge t'\cos 2 k_x-t \cos k_y-\mu\ge -t'-t-\mu\ ,
\end{equation}
We also check that for $\bq=(\pm\pi/2,0,q_z)$ state in lower band:
\begin{equation}\tag{S4}
\epsilon_{\bq,-}=-t'-t-|\lambda_z|\ ,
\end{equation}
If chemical potential $\mu$ satisfies $-t'-t-|\lambda_z|<\mu<-t'-t$, we have $\epsilon_{\bk,+}>0$ for all $\bk$ which indicates that all the upper band states are unoccupied and $\epsilon_{\bq,-}<0$ which indicates that the states in the lower band near $(k_x,k_y)=(\pm \pi/2,0)$ are occupied. The fermi surfaces are therefore two cylinders at $(k_x,k_y)=(\pm \pi/2,0)$ shown as Fig.1a. For higher chemical potential, the fermi surfaces can have both inner and outer cylinders at $(k_x,k_y)=(\pm\pi/2,0)$. In this article, we consider band and SOC parameters when the fermi surfaces are a pair of cylinders as Fig.1a.

\subsection{Projected interaction on the fermi surface} \label{projected-interaction}
In this supplemental subsection, we shall follow the single particle Hamiltonian with spiral spin texture Eq.(4) and interaction Eq.(5) to show that effective interaction term such as Eq.(1) can be realized. The single particle Hamiltonian has two cylindrical fermi surfaces at $(k_x,k_y)=(\pm \pi/2,0,0)$ as shown in Fig.3(a).  We now calculate the projected interaction on fermi surfaces. For a superconductor,  we assume that Cooper channel dominates the interaction effect and neglects other interaction terms. It is a good approximation to keep only the Cooper channel components of electron-electron interaction:
\begin{equation}\tag{S5}
H_{\text{Cooper}}=\frac{1}{2 N}\sum_{\bk,\bk'}V_{s_{1},s_{2}, s_{2}', s_{1}'}(\bk,\bk') c_{\bk s_{1}}^{\dagger} c_{-\bk s_{2}}^{\dagger} c_{-\bk' s_{2}'} c_{\bk' s_{1}'},
\end{equation}
with
\begin{equation}\tag{S6}
\begin{split}
V_{s_{1},s_{2}, s_{2}', s_{1}'}(\bk,\bk')&=V(\bk'-\bk)\delta_{s_{1} s_{1}'}\delta_{s_{2} s_{2}'}-V(\bk'+\bk)\delta_{s_{2} s_{1}'}\delta_{s_{1} s_{2}'}\\
&+\frac{1}{4}J(\bk'-\bk)\bsigma_{s_{1} s_{1}'}\cdot \bsigma_{s_{2} s_{2}'}-\frac{1}{4}J(\bk'+\bk)\bsigma_{s_{2} s_{1}'}\cdot \bsigma_{s_{1} s_{2}'},
\end{split}
\label{Cooper}
\end{equation}
We first discuss the case where $\bk$ and $\bk'$ locate on the cylindrical fermi surface at $k_x=\pi/2$. In our lattice model Eq.(4), the spin texture is characterized by $\phi(\bk)$, the angle of spin direction at $\bk$ in the $k_x-k_y$ plane. At momentum $\bk$, the spin wavefunction is $\frac{1}{\sqrt{2}}\left(1, e^{i\phi(\bk)}\right)^{T}$ while the time reversal state at the other cylinder is $\frac{1}{\sqrt{2}}\left( e^{-i\phi(\bk)},-1\right)^{T}$. This choice of basis will keep pairing function real as long as the ground state preserves time reversal symmetry\cite{qi2009}. We can now project the Cooper channel onto fermi surfaces using the spin states wave function. The projected interaction $V_{\bk\bk'}$ is obtained as
\begin{equation}\tag{S7}
\begin{split}
V_{\bk \bk'}&=\frac{1}{2}\left[V(\bk'-\bk)+V(\bk'+\bk)-\frac{3}{4}J(\bk'-\bk)-\frac{3}{4}J(\bk'+\bk)\right]\\
&+\frac{1}{2}\left[V(\bk'-\bk)-V(\bk'+\bk)+\frac{1}{4}J(\bk'-\bk)-\frac{1}{4}J(\bk'+\bk)\right]\cos[\phi(\bk)-\phi(\bk')],
\end{split}
\end{equation}
The definition of $V_{\bk\bk'}$ is the same as the main text in which we write the projected interaction of form $\sum_{\bk\bk'}(V_{\bk\bk'}/2N) c_{\bk}^{\dagger}c_{-\bk}^{\dagger} c_{-\bk'}c_{\bk'}$. Here $c_{\bk}$ is fermion operator of momentum $\bk$ defined on the projected band. For specific interaction potentials mentioned in main text, $V(\bq)=V_0+V_x \cos 2 k_x+V_y\cos k_y$ and $J(\bq)=J_x \cos 2 k_x+J_y \cos k_y$,
the corresponding projected interaction for momentums $\bk$ and $\bk'$ near the cylindrical fermi surface can be written as:
\begin{equation}\tag{S8}
\begin{split}
V_{\bk\bk'}&\approx  \left(V_{0}+V_{x}-\frac{3}{4}J_{x}+V_y-\frac{3}{4}J_y\right)-2 \left(V_{x}-\frac{3}{4}J_{x}\right)(\widetilde{k_{x}}^2+\widetilde{k_{x}'}^2)-\frac{1}{2}\left(V_y-\frac{3}{4}J_y\right)(k_y^2+k_y'^2)\\
&+ 4 \left(V_{x}+\frac{1}{4}J_{x}\right) \widetilde{k_{x}}\widetilde{k_{x}'} \cos[\phi(\bk)-\phi(\bk')]+\left(V_y+\frac{1}{4}J_y \right)k_{y}k_{y}'\cos[\phi(\bk)-\phi(\bk')],
\end{split}
\end{equation}
where we have considered cylindrical fermi surfaces to be small and expanded up to the second order of $\widetilde{k_{x}}$ and $k_y$. We shall keep this approximation in the following subsections. The projected interaction of other momentums can be obtained from the relation: $V_{\bk\bk'}=-V_{\bk,-\bk'}=-V_{-\bk,\bk'}$. A rearrangement of variables will give the result of Eq.(6) in the main text.

\subsection{Phase diagram for small $\lambda_z/\lambda$}\label{small-kz-dependent}
In this supplemental subsection, we shall discuss the phase diagram for small $\lambda_z/\lambda$. We adopt projected interaction Eq.(S8) and solve numerically the linearized gap equation Eq.(7). We fix single-particle parameters at $t'=0.25t$, $\lambda=-0.2t$, $\lambda_z=0.08 t$ when $\lambda_z/\lambda$ is relatively small. As a comparison of Fig.3(a), we compute phase diagram {\textbf{Supplemental figure}} 1{\textbf{a}} for $J_x=J_y=J$ and $V_0$ with fixed $V_x=V_y=V<0$. 
\begin{figure*}
\includegraphics[width=0.65\textwidth]{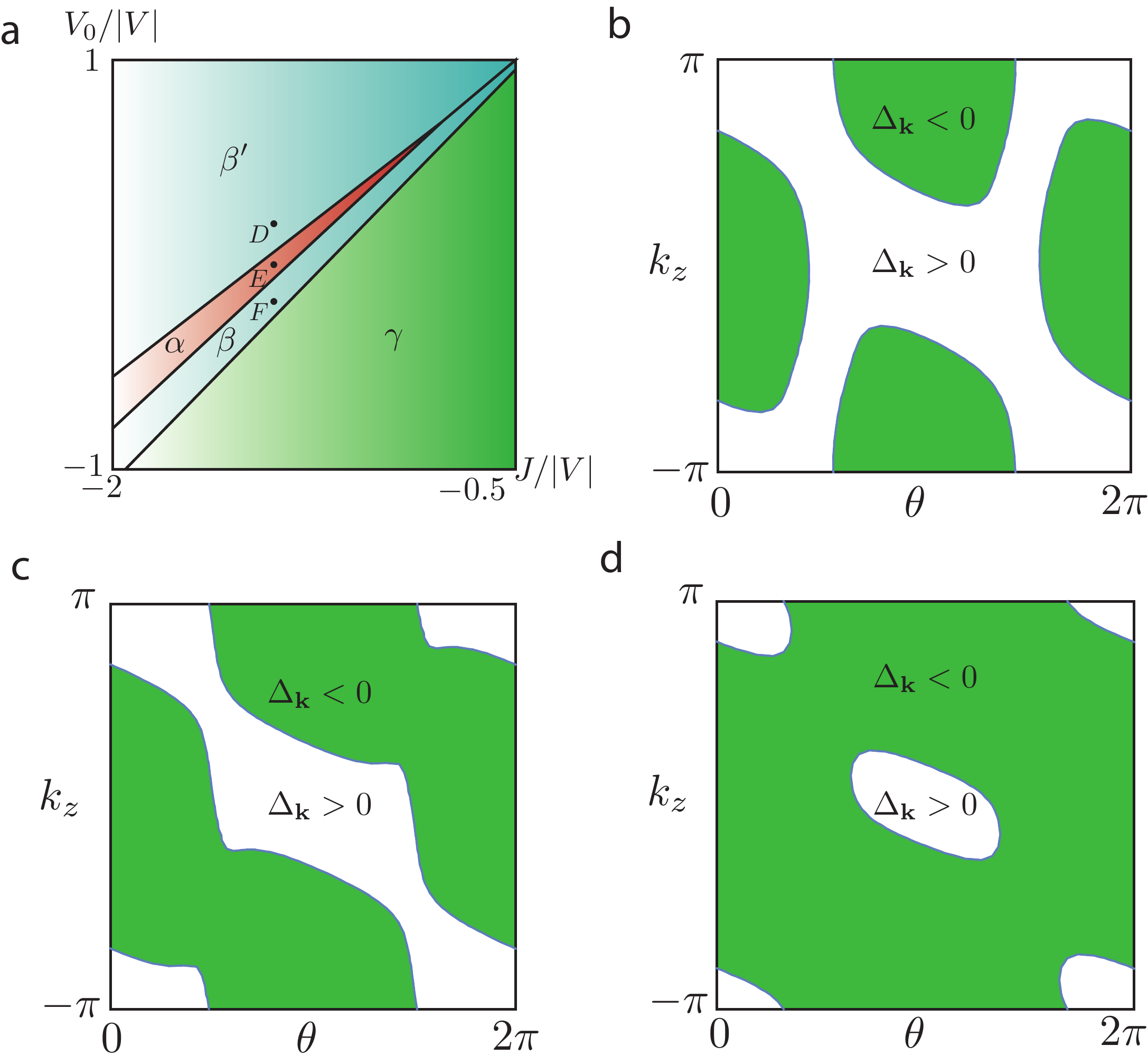}
\caption{{\textbf{a}}.\ Phase diagram for fixed $V<0$ for small $k_z$ dependent SOC, where the horizontal axis and the vertical axis are $J$ and $V_0$, respectively.  $\alpha$ phase is the double-helix linked nodal line phase. The $\beta$ and $\beta'$ phases contain two unlinked nodal lines encircling at each fermi surface. Nodal lines encircles different $TM_x$ invariant momentums in $\beta$ and $\beta'$ phases. The $\gamma$ phase is the conventional fully gapped phase. Panels {\textbf{b-d}} show the nodal line configuration on the fermi surface at $k_x=\pi/2$ for points $D$, $E$ and $F$ in the phase diagram, respectively.}
\label{phase-diagram-supplemental}
\end{figure*}
Our lattice model Eq.(4) possesses time-reversal symmetry $T$ and mirror $x$ symmetry $M_x: x\rightarrow-x$. The combination symmetry $TM_x$ transforms each of the cylindrical fermi surface to itself. We parametrize torus fermi surface located at $k_x=\pi/2$ by $k_z$ and poloidal angle $\theta=\arg(\widetilde{k_x}+i k_y)$. On fermi surface, there are four $TM_x$ invariant momentums located at $(\theta, k_z)=(0,0)$, $(0,\pi)$, $(\pi, 0)$ and $(\pi,\pi)$. $T M_x$ invariant nodal loop encircling $TM_x$ invariant momentum is non-trivial. As long as the ground state preserves $TM_x$ symmetry, one can not tune parameters to move the center of such $T M_x$ invariant nodal loop to other momentum. Feasible ways of removing $TM_x$ invariant loop is merging with other nodal loop or shrinking it to $TM_x$ invariant momentum. Each of them will accompany with a Lifshitz transition. In the phase diagram, both $\beta$ and $\beta'$ phases have two unlinked nodal loops. $\beta$ phase has two nodal loops at $(\theta, k_z)=(0,\pi)$ and $(\pi,0)$ while $\beta'$ phase has two nodal loops at $(\theta, k_z)=(0,0)$ and $(\pi,\pi)$ as shown in {\textbf{Supplemental figure}} 1{\textbf{b}} and 1{\textbf{d}}. The two unlinked nodal loop phases are protected by $TM_x$ symmetry. The $\alpha$ phase is the only double-helix nodal-line phase with a left-handed double helix that lies between the phase space of $\beta$ phase and $\beta'$ phase. We may understand the phase diagram with fixed ferromagnetic interaction $J/|V|$ by applying physical picture of Fig.2. As $V_0/|V|$ decreases starting with $\beta'$ phase, attractive $V_0$ favors singlet pairing and singlet pairing order parameter $\psi$ becomes stronger. As shown in Fig.2, near $k_z=0$ plane, $|\bd(\bk)|=\psi/\gamma$ contour will expand and consequently enlarge nodal loop at $(\theta, k_z)=(0,0)$. Similarly, nodal loop at $(\theta, k_z)=(\pi,\pi)$ also expand as $V_0/|V|$ decreases. At a critical value of $V_0/|V|$, two nodal loops merge and reconnect as double helix nodal lines in phase $\alpha$ as shown in {\textbf{Supplemental figure}} 1{\textbf{c}}. Similar argument applies if we start from $\beta$ phase and increase $V_0/|V|$. Nodal loops at $(\theta, k_z)=(0,\pi)$ and $(\pi,0)$ will expand. At a critical value of $V_0/|V|$, two nodal loops merge and reconnect as double helix nodal lines in phase $\alpha$. At the end of this supplemental subsection, we remark on $\alpha'$ phase which is the difference of phase diagram {\textbf{Supplemental figure}} 1{\textbf{a}} and Fig.3(a). In contrary to $\alpha$ phase, at fermi surface located at $k_x=\pi/2$, $\alpha'$ phase is right-handed with the opposite chirality of spiral spin texture. The physical picture of Fig.2 predicts the chirality of double helix to be the same with that of spiral spin texture and only the $\alpha$ phase is predicted. Based on all the observations, the numerical calculation supports physical picture for small $\lambda_z/\lambda$ in the main text.

\subsection{Double helix nodal line phase at zero temperature}\label{zero-temperature}
In the main text, we show the phase diagram of the highest $T_c$ pairing function with different interactions based on nodal lines topology. This supplemental subsection will show that the phase diagram defined by topology will not have qualitative variation down to zero temperature. We shall start from the general self-consistent gap equation for superconductivity:
\begin{equation}\tag{S5}
\Delta_{\bk}=-\frac{1}{N}\sum_{\bk'}V_{\bk\bk'}\frac{\Delta_{\bk'}}{2E_{\bk'}}[1-2 n_{F}(E_{\bk'})], 
\end{equation}
For implementation in numerics, we discretize $\bk$ and $\bk'$ momentums in the energy shell ranges from $-\epsilon_c$ to $+\epsilon_c$. In the limit of $T\rightarrow T_c$, we can expand both sides for small $\Delta(\bk)$ and derive the linear gap equation for $T_c$:
\begin{equation}\tag{S6}
-\chi \Psi(\bk)=\frac{1}{2N\epsilon_c}\sum_{\bk'} V_{\bk \bk'} \Psi(\bk')\ ,
\end{equation} 
where $\Psi(\bk)$ is proportional to $\Delta(\bk)$ near $T_c$. This is an eigenvalue equation that can be solved by numerically diagonalize matrix $V_{\bk \bk'}$, the largest positive $\chi$ represents superconducting instability and determines critical temperature $T_c$. The phase diagram by solving linear gap equation is presented in Fig3.b. Here we consider the other limit of zero temperature, the gap equation is:
\begin{equation}\tag{S7}
\Delta_{\bk}=-\frac{1}{N}\sum_{\bk'} V_{\bk\bk'}\frac{\Delta_{\bk'}}{2E_{\bk'}}, 
\label{gap-equation-zero}
\end{equation}
We solve this non-linear equation by iteration. Generically, we find that the linking number between nodal lines do not change from $T_c$ to zero temperature. As a specific example, we show the geometry of nodal lines in {\textbf{Supplemental figure}} 2{\textbf{a}} with the same parameter as Fig.3(a) in the main text.

\subsection{Stability of double helix nodal line phase under symmetry allowed perturbation}\label{spin-canting}
In our lattice model, we assume a perfect spiral spin texture with spin polarized in $k_x-k_y$ plane and a specific pattern of hopping parameters. Linking number, however, with a topological nature, should not change under small perturbation that preserves time-reversal symmetry. Here we present an example perturbation of a spin canting term in our lattice model:
\begin{equation}\tag{S8}
H'=\lambda_5 \sin k_x \sigma_z.
\end{equation}
With the same strategy as subsection \ref{projected-interaction} and \ref{zero-temperature}, we project electron-electron interaction on two cylindrical torus fermi surfaces and solve gap equation. As a comparison, we perturb the Hamiltonian in Fig.3(a) in the main text by $H'$ with $\lambda_5=0.2t$ and show the nodal lines in in {\textbf{Supplemental figure}} 2{\textbf{b}}. Generically, one may check that symmetry allowed perturbation terms will not change the linking number of nodal lines. 
\begin{figure*}
\includegraphics[width=0.9\textwidth]{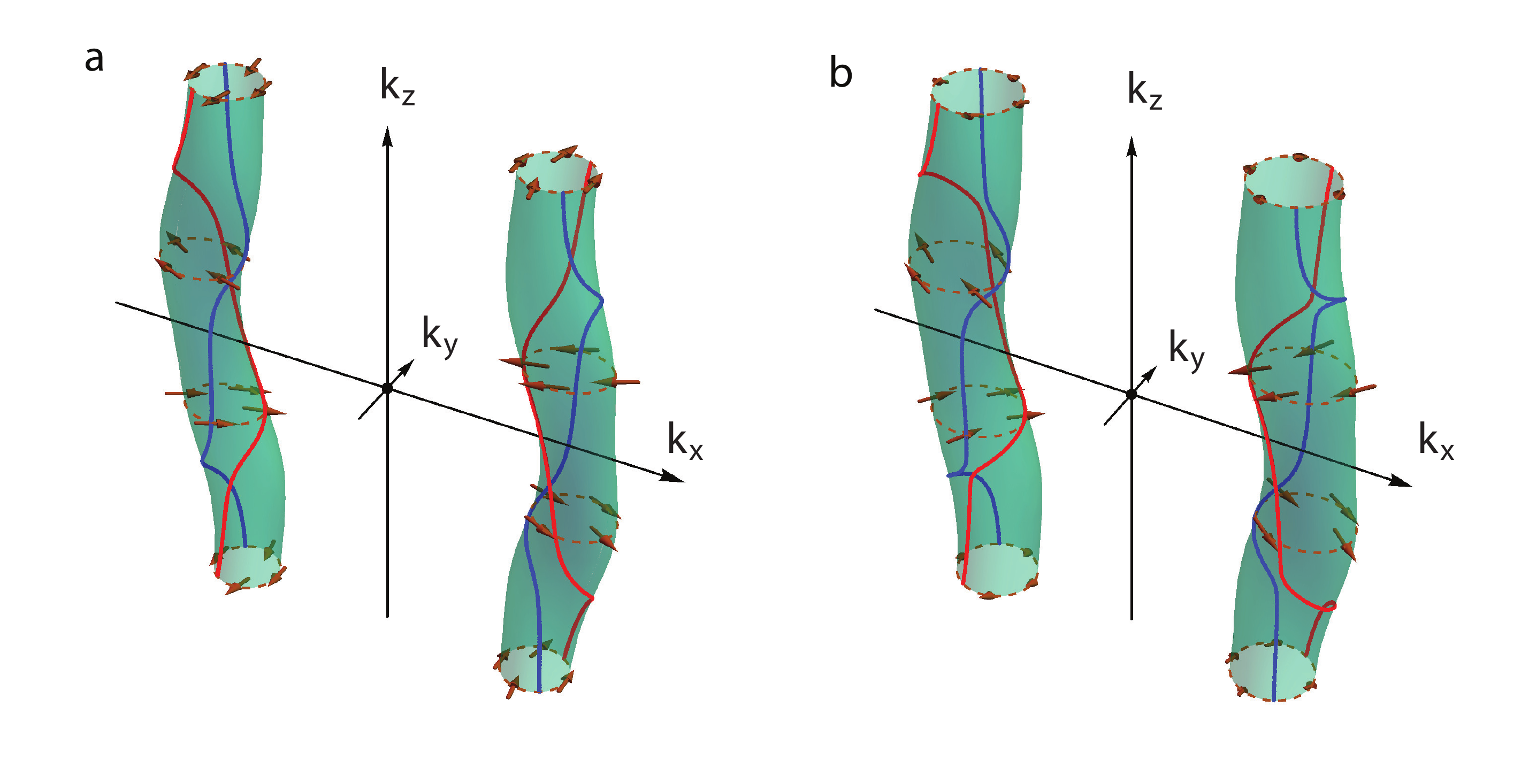}
\label{suppfig}
\caption{{\textbf{a}}.\ The plot of spin texture and nodal lines on fermi surface at zero temperature of same Hamiltonian as in Fig.3(a).  {\textbf{b}}.\ The plot of spin texture and nodal lines on fermi surface after adding $H'$($\lambda_5=0.2t'$) to Hamiltonian in Fig.3(a).}
\end{figure*}
\end{widetext}


\begin{thebibliography}{35}
\expandafter\ifx\csname natexlab\endcsname\relax\def\natexlab#1{#1}\fi
\expandafter\ifx\csname bibnamefont\endcsname\relax
  \def\bibnamefont#1{#1}\fi
\expandafter\ifx\csname bibfnamefont\endcsname\relax
  \def\bibfnamefont#1{#1}\fi
\expandafter\ifx\csname citenamefont\endcsname\relax
  \def\citenamefont#1{#1}\fi
\expandafter\ifx\csname url\endcsname\relax
  \def\url#1{\texttt{#1}}\fi
\expandafter\ifx\csname urlprefix\endcsname\relax\def\urlprefix{URL }\fi
\providecommand{\bibinfo}[2]{#2}
\providecommand{\eprint}[2][]{\url{#2}}

\bibitem[{\citenamefont{Hu}(1994)}]{hu1994}
\bibinfo{author}{\bibfnamefont{C.-R.} \bibnamefont{Hu}},
  \bibinfo{journal}{Phys. Rev. Lett.} \textbf{\bibinfo{volume}{72}},
  \bibinfo{pages}{1526} (\bibinfo{year}{1994}).

\bibitem[{\citenamefont{Wollman et~al.}(1993)\citenamefont{Wollman,
  Van~Harlingen, Lee, Ginsberg, and Leggett}}]{wollman1993}
\bibinfo{author}{\bibfnamefont{D.}~\bibnamefont{Wollman}},
  \bibinfo{author}{\bibfnamefont{D.}~\bibnamefont{Van~Harlingen}},
  \bibinfo{author}{\bibfnamefont{W.}~\bibnamefont{Lee}},
  \bibinfo{author}{\bibfnamefont{D.}~\bibnamefont{Ginsberg}}, \bibnamefont{and}
  \bibinfo{author}{\bibfnamefont{A.}~\bibnamefont{Leggett}},
  \bibinfo{journal}{Phys. Rev. Lett.} \textbf{\bibinfo{volume}{71}},
  \bibinfo{pages}{2134} (\bibinfo{year}{1993}).

\bibitem[{\citenamefont{Wollman et~al.}(1995)\citenamefont{Wollman,
  Van~Harlingen, Giapintzakis, and Ginsberg}}]{wollman1995}
\bibinfo{author}{\bibfnamefont{D.}~\bibnamefont{Wollman}},
  \bibinfo{author}{\bibfnamefont{D.}~\bibnamefont{Van~Harlingen}},
  \bibinfo{author}{\bibfnamefont{J.}~\bibnamefont{Giapintzakis}},
  \bibnamefont{and} \bibinfo{author}{\bibfnamefont{D.}~\bibnamefont{Ginsberg}},
  \bibinfo{journal}{Phys. Rev. Lett.} \textbf{\bibinfo{volume}{74}},
  \bibinfo{pages}{797} (\bibinfo{year}{1995}).

\bibitem[{\citenamefont{Kirtley et~al.}(1995)\citenamefont{Kirtley, Tsuei, Sun,
  Chi et~al.}}]{kirtley1995}
\bibinfo{author}{\bibfnamefont{J.}~\bibnamefont{Kirtley}},
  \bibinfo{author}{\bibfnamefont{C.}~\bibnamefont{Tsuei}},
  \bibinfo{author}{\bibfnamefont{J.}~\bibnamefont{Sun}},
  \bibinfo{author}{\bibfnamefont{C.}~\bibnamefont{Chi}}, \bibnamefont{et~al.},
  \bibinfo{journal}{Nature (London)} \textbf{\bibinfo{volume}{373}},
  \bibinfo{pages}{225} (\bibinfo{year}{1995}).

\bibitem[{\citenamefont{Okazaki et~al.}(2012)\citenamefont{Okazaki, Ota,
  Kotani, Malaeb, Ishida, Shimojima, Kiss, Watanabe, Chen, Kihou
  et~al.}}]{okazaki2012}
\bibinfo{author}{\bibfnamefont{K.}~\bibnamefont{Okazaki}},
  \bibinfo{author}{\bibfnamefont{Y.}~\bibnamefont{Ota}},
  \bibinfo{author}{\bibfnamefont{Y.}~\bibnamefont{Kotani}},
  \bibinfo{author}{\bibfnamefont{W.}~\bibnamefont{Malaeb}},
  \bibinfo{author}{\bibfnamefont{Y.}~\bibnamefont{Ishida}},
  \bibinfo{author}{\bibfnamefont{T.}~\bibnamefont{Shimojima}},
  \bibinfo{author}{\bibfnamefont{T.}~\bibnamefont{Kiss}},
  \bibinfo{author}{\bibfnamefont{S.}~\bibnamefont{Watanabe}},
  \bibinfo{author}{\bibfnamefont{C.-T.} \bibnamefont{Chen}},
  \bibinfo{author}{\bibfnamefont{K.}~\bibnamefont{Kihou}},
  \bibnamefont{et~al.}, \bibinfo{journal}{Science}
  \textbf{\bibinfo{volume}{337}}, \bibinfo{pages}{1314} (\bibinfo{year}{2012}).

\bibitem[{\citenamefont{Zhang et~al.}(2012)\citenamefont{Zhang, Ye, Ge, Chen,
  Jiang, Xu, Xie, and Feng}}]{zhang2012}
\bibinfo{author}{\bibfnamefont{Y.}~\bibnamefont{Zhang}},
  \bibinfo{author}{\bibfnamefont{Z.}~\bibnamefont{Ye}},
  \bibinfo{author}{\bibfnamefont{Q.}~\bibnamefont{Ge}},
  \bibinfo{author}{\bibfnamefont{F.}~\bibnamefont{Chen}},
  \bibinfo{author}{\bibfnamefont{J.}~\bibnamefont{Jiang}},
  \bibinfo{author}{\bibfnamefont{M.}~\bibnamefont{Xu}},
  \bibinfo{author}{\bibfnamefont{B.}~\bibnamefont{Xie}}, \bibnamefont{and}
  \bibinfo{author}{\bibfnamefont{D.}~\bibnamefont{Feng}},
  \bibinfo{journal}{Nat. Phys.} \textbf{\bibinfo{volume}{8}},
  \bibinfo{pages}{371} (\bibinfo{year}{2012}).

\bibitem[{\citenamefont{Bauer et~al.}(2004)\citenamefont{Bauer, Hilscher,
  Michor, Paul, Scheidt, Gribanov, Seropegin, No{\"e}l, Sigrist, and
  Rogl}}]{bauer2004}
\bibinfo{author}{\bibfnamefont{E.}~\bibnamefont{Bauer}},
  \bibinfo{author}{\bibfnamefont{G.}~\bibnamefont{Hilscher}},
  \bibinfo{author}{\bibfnamefont{H.}~\bibnamefont{Michor}},
  \bibinfo{author}{\bibfnamefont{C.}~\bibnamefont{Paul}},
  \bibinfo{author}{\bibfnamefont{E.}~\bibnamefont{Scheidt}},
  \bibinfo{author}{\bibfnamefont{A.}~\bibnamefont{Gribanov}},
  \bibinfo{author}{\bibfnamefont{Y.}~\bibnamefont{Seropegin}},
  \bibinfo{author}{\bibfnamefont{H.}~\bibnamefont{No{\"e}l}},
  \bibinfo{author}{\bibfnamefont{M.}~\bibnamefont{Sigrist}}, \bibnamefont{and}
  \bibinfo{author}{\bibfnamefont{P.}~\bibnamefont{Rogl}},
  \bibinfo{journal}{Phys. Rev. Lett.} \textbf{\bibinfo{volume}{92}},
  \bibinfo{pages}{027003} (\bibinfo{year}{2004}).

\bibitem[{\citenamefont{Yip}(2014)}]{yip2014}
\bibinfo{author}{\bibfnamefont{S.}~\bibnamefont{Yip}}, \bibinfo{journal}{Annu.
  Rev. Condens. Matter Phys.} \textbf{\bibinfo{volume}{5}}, \bibinfo{pages}{15}
  (\bibinfo{year}{2014}).

\bibitem[{\citenamefont{Lian et~al.}(2017)\citenamefont{Lian, Vafa, Vafa, and
  Zhang}}]{lian2017}
\bibinfo{author}{\bibfnamefont{B.}~\bibnamefont{Lian}},
  \bibinfo{author}{\bibfnamefont{C.}~\bibnamefont{Vafa}},
  \bibinfo{author}{\bibfnamefont{F.}~\bibnamefont{Vafa}}, \bibnamefont{and}
  \bibinfo{author}{\bibfnamefont{S.-C.} \bibnamefont{Zhang}},
  \bibinfo{journal}{Phys. Rev. B} \textbf{\bibinfo{volume}{95}},
  \bibinfo{pages}{094512} (\bibinfo{year}{2017}).

\bibitem[{\citenamefont{Chen et~al.}(2017)\citenamefont{Chen, Lu, and
  Hou}}]{chen2017}
\bibinfo{author}{\bibfnamefont{W.}~\bibnamefont{Chen}},
  \bibinfo{author}{\bibfnamefont{H.-Z.} \bibnamefont{Lu}}, \bibnamefont{and}
  \bibinfo{author}{\bibfnamefont{J.-M.} \bibnamefont{Hou}},
  \bibinfo{journal}{Phys. Rev. B} \textbf{\bibinfo{volume}{96}},
  \bibinfo{pages}{041102} (\bibinfo{year}{2017}).

\bibitem[{\citenamefont{Yan et~al.}(2017)\citenamefont{Yan, Bi, Shen, Lu,
  Zhang, and Wang}}]{yan2017}
\bibinfo{author}{\bibfnamefont{Z.}~\bibnamefont{Yan}},
  \bibinfo{author}{\bibfnamefont{R.}~\bibnamefont{Bi}},
  \bibinfo{author}{\bibfnamefont{H.}~\bibnamefont{Shen}},
  \bibinfo{author}{\bibfnamefont{L.}~\bibnamefont{Lu}},
  \bibinfo{author}{\bibfnamefont{S.-C.} \bibnamefont{Zhang}}, \bibnamefont{and}
  \bibinfo{author}{\bibfnamefont{Z.}~\bibnamefont{Wang}},
  \bibinfo{journal}{Phys. Rev. B} \textbf{\bibinfo{volume}{96}},
  \bibinfo{pages}{041103} (\bibinfo{year}{2017}).

\bibitem[{\citenamefont{Ezawa}(2017)}]{ezawa2017}
\bibinfo{author}{\bibfnamefont{M.}~\bibnamefont{Ezawa}},
  \bibinfo{journal}{Phys. Rev. B} \textbf{\bibinfo{volume}{96}},
  \bibinfo{pages}{041202} (\bibinfo{year}{2017}).

\bibitem[{\citenamefont{Wang et~al.}(2011)\citenamefont{Wang, Qi, and
  Zhang}}]{wang2011}
\bibinfo{author}{\bibfnamefont{Z.}~\bibnamefont{Wang}},
  \bibinfo{author}{\bibfnamefont{X.-L.} \bibnamefont{Qi}}, \bibnamefont{and}
  \bibinfo{author}{\bibfnamefont{S.-C.} \bibnamefont{Zhang}},
  \bibinfo{journal}{Phys. Rev. B} \textbf{\bibinfo{volume}{84}},
  \bibinfo{pages}{014527} (\bibinfo{year}{2011}).

\bibitem[{\citenamefont{Ryu et~al.}(2012)\citenamefont{Ryu, Moore, and
  Ludwig}}]{ryu2012}
\bibinfo{author}{\bibfnamefont{S.}~\bibnamefont{Ryu}},
  \bibinfo{author}{\bibfnamefont{J.~E.} \bibnamefont{Moore}}, \bibnamefont{and}
  \bibinfo{author}{\bibfnamefont{A.~W.~W.} \bibnamefont{Ludwig}},
  \bibinfo{journal}{Phys. Rev. B} \textbf{\bibinfo{volume}{85}},
  \bibinfo{pages}{045104} (\bibinfo{year}{2012}).

\bibitem[{\citenamefont{Polyakov}(1988)}]{polyakov1988}
\bibinfo{author}{\bibfnamefont{A.~M.} \bibnamefont{Polyakov}},
  \bibinfo{journal}{Mod. Phys. Lett. A} \textbf{\bibinfo{volume}{3}},
  \bibinfo{pages}{325} (\bibinfo{year}{1988}).

\bibitem[{\citenamefont{Witten}(1989)}]{witten1989}
\bibinfo{author}{\bibfnamefont{E.}~\bibnamefont{Witten}},
  \bibinfo{journal}{Commun. Math. Phys.} \textbf{\bibinfo{volume}{121}},
  \bibinfo{pages}{351} (\bibinfo{year}{1989}).

\bibitem[{\citenamefont{Schnyder and Ryu}(2011)}]{schnyder2011}
\bibinfo{author}{\bibfnamefont{A.~P.} \bibnamefont{Schnyder}} \bibnamefont{and}
  \bibinfo{author}{\bibfnamefont{S.}~\bibnamefont{Ryu}},
  \bibinfo{journal}{Phys. Rev. B} \textbf{\bibinfo{volume}{84}},
  \bibinfo{pages}{060504} (\bibinfo{year}{2011}).

\bibitem[{\citenamefont{Schnyder et~al.}(2012)\citenamefont{Schnyder, Brydon,
  and Timm}}]{schnyder2012}
\bibinfo{author}{\bibfnamefont{A.~P.} \bibnamefont{Schnyder}},
  \bibinfo{author}{\bibfnamefont{P.}~\bibnamefont{Brydon}}, \bibnamefont{and}
  \bibinfo{author}{\bibfnamefont{C.}~\bibnamefont{Timm}},
  \bibinfo{journal}{Phys. Rev. B} \textbf{\bibinfo{volume}{85}},
  \bibinfo{pages}{024522} (\bibinfo{year}{2012}).

\bibitem[{\citenamefont{Schnyder and Brydon}(2015)}]{schnyder2015}
\bibinfo{author}{\bibfnamefont{A.~P.} \bibnamefont{Schnyder}} \bibnamefont{and}
  \bibinfo{author}{\bibfnamefont{P.~M.} \bibnamefont{Brydon}},
  \bibinfo{journal}{J. Phys. Condensed Matter} \textbf{\bibinfo{volume}{27}},
  \bibinfo{pages}{243201} (\bibinfo{year}{2015}).

\bibitem[{\citenamefont{Sato et~al.}(2011)\citenamefont{Sato, Tanaka, Yada, and
  Yokoyama}}]{sato2011}
\bibinfo{author}{\bibfnamefont{M.}~\bibnamefont{Sato}},
  \bibinfo{author}{\bibfnamefont{Y.}~\bibnamefont{Tanaka}},
  \bibinfo{author}{\bibfnamefont{K.}~\bibnamefont{Yada}}, \bibnamefont{and}
  \bibinfo{author}{\bibfnamefont{T.}~\bibnamefont{Yokoyama}},
  \bibinfo{journal}{Phys. Rev. B} \textbf{\bibinfo{volume}{83}},
  \bibinfo{pages}{224511} (\bibinfo{year}{2011}).

\bibitem[{\citenamefont{Tanaka et~al.}(2011)\citenamefont{Tanaka, Sato, and
  Nagaosa}}]{tanaka2011}
\bibinfo{author}{\bibfnamefont{Y.}~\bibnamefont{Tanaka}},
  \bibinfo{author}{\bibfnamefont{M.}~\bibnamefont{Sato}}, \bibnamefont{and}
  \bibinfo{author}{\bibfnamefont{N.}~\bibnamefont{Nagaosa}},
  \bibinfo{journal}{J. Phys. Soc. Jpn} \textbf{\bibinfo{volume}{81}},
  \bibinfo{pages}{011013} (\bibinfo{year}{2011}).

\bibitem[{\citenamefont{Watson and Crick}(1953)}]{watson1953}
\bibinfo{author}{\bibfnamefont{J.~D.} \bibnamefont{Watson}} \bibnamefont{and}
  \bibinfo{author}{\bibfnamefont{F.~H.} \bibnamefont{Crick}},
  \bibinfo{journal}{Nature (London)} \textbf{\bibinfo{volume}{171}},
  \bibinfo{pages}{737} (\bibinfo{year}{1953}).

\bibitem[{\citenamefont{Watson}(2012)}]{watson2012}
\bibinfo{author}{\bibfnamefont{J.}~\bibnamefont{Watson}},
  \emph{\bibinfo{title}{The Double Helix}} (\bibinfo{publisher}{Hachette UK},
  \bibinfo{year}{2012}).

\bibitem[{\citenamefont{Anderson}(1984)}]{anderson1984}
\bibinfo{author}{\bibfnamefont{P.}~\bibnamefont{Anderson}},
  \bibinfo{journal}{Phys. Rev. B} \textbf{\bibinfo{volume}{30}},
  \bibinfo{pages}{4000} (\bibinfo{year}{1984}).

\bibitem[{\citenamefont{Qi et~al.}(2009)\citenamefont{Qi, Hughes, Raghu, and
  Zhang}}]{qi2009}
\bibinfo{author}{\bibfnamefont{X.-L.} \bibnamefont{Qi}},
  \bibinfo{author}{\bibfnamefont{T.~L.} \bibnamefont{Hughes}},
  \bibinfo{author}{\bibfnamefont{S.}~\bibnamefont{Raghu}}, \bibnamefont{and}
  \bibinfo{author}{\bibfnamefont{S.-C.} \bibnamefont{Zhang}},
  \bibinfo{journal}{Phys. Rev. Lett.} \textbf{\bibinfo{volume}{102}},
  \bibinfo{pages}{187001} (\bibinfo{year}{2009}).

\bibitem[{\citenamefont{Gauss}()}]{gauss1867}
\bibinfo{author}{\bibfnamefont{C. F. }~\bibnamefont{Gauss}},
  \emph{\bibinfo{title}{\textit{Werke}, (Kr{\""o}nigliche Gesellschaft der Wissenschaften, G{\"o}ttingen, 1833), Vol. 5, p. 605}}.

\bibitem[{\citenamefont{Sigrist and Ueda}(1991)}]{sigrist1991}
\bibinfo{author}{\bibfnamefont{M.}~\bibnamefont{Sigrist}} \bibnamefont{and}
  \bibinfo{author}{\bibfnamefont{K.}~\bibnamefont{Ueda}},
  \bibinfo{journal}{Rev. Mod. Phys.} \textbf{\bibinfo{volume}{63}},
  \bibinfo{pages}{239} (\bibinfo{year}{1991}).

\bibitem[{\citenamefont{Frigeri et~al.}(2004)\citenamefont{Frigeri, Agterberg,
  Koga, and Sigrist}}]{frigeri2004}
\bibinfo{author}{\bibfnamefont{P.}~\bibnamefont{Frigeri}},
  \bibinfo{author}{\bibfnamefont{D.}~\bibnamefont{Agterberg}},
  \bibinfo{author}{\bibfnamefont{A.}~\bibnamefont{Koga}}, \bibnamefont{and}
  \bibinfo{author}{\bibfnamefont{M.}~\bibnamefont{Sigrist}},
  \bibinfo{journal}{Phys. Rev. Lett.} \textbf{\bibinfo{volume}{92}},
  \bibinfo{pages}{097001} (\bibinfo{year}{2004}).

\bibitem[{sup()}]{suppl}
\bibinfo{note}{See Supplemental Material for technical details on (i) the Fermi surface of single particle Hamiltonian, (ii) the projected interaction on the Fermi surface, (iii) the phase diagram for small $\lambda_z/\lambda$, (iv) the double helix nodal-line phase at zero temperature, (v) and the stability of the double helix nodal-line phase under symmetry allowed perturbations.}

\bibitem[{\citenamefont{Sigrist}(2005)}]{sigrist2005}
\bibinfo{author}{\bibfnamefont{M.}~\bibnamefont{Sigrist}},
  \bibinfo{journal}{AIP Conf. Proc.} \textbf{\bibinfo{volume}{789}},
  \bibinfo{pages}{165} (\bibinfo{year}{2005}).

\bibitem[{\citenamefont{Chhowalla et~al.}(2013)\citenamefont{Chhowalla, Shin,
  Eda, Li, Loh, and Zhang}}]{chhowalla2013}
\bibinfo{author}{\bibfnamefont{M.}~\bibnamefont{Chhowalla}},
  \bibinfo{author}{\bibfnamefont{H.~S.} \bibnamefont{Shin}},
  \bibinfo{author}{\bibfnamefont{G.}~\bibnamefont{Eda}},
  \bibinfo{author}{\bibfnamefont{L.-J.} \bibnamefont{Li}},
  \bibinfo{author}{\bibfnamefont{K.~P.} \bibnamefont{Loh}}, \bibnamefont{and}
  \bibinfo{author}{\bibfnamefont{H.}~\bibnamefont{Zhang}},
  \bibinfo{journal}{Nat. Chem.} \textbf{\bibinfo{volume}{5}},
  \bibinfo{pages}{263} (\bibinfo{year}{2013}).

\bibitem[{\citenamefont{Mir{\'o} et~al.}(2014)\citenamefont{Mir{\'o},
  Audiffred, and Heine}}]{miro2014}
\bibinfo{author}{\bibfnamefont{P.}~\bibnamefont{Mir{\'o}}},
  \bibinfo{author}{\bibfnamefont{M.}~\bibnamefont{Audiffred}},
  \bibnamefont{and} \bibinfo{author}{\bibfnamefont{T.}~\bibnamefont{Heine}},
  \bibinfo{journal}{Chem. Soc. Rev.} \textbf{\bibinfo{volume}{43}},
  \bibinfo{pages}{6537} (\bibinfo{year}{2014}).

\bibitem[{\citenamefont{Van~Harlingen}(1995)}]{van1995}
\bibinfo{author}{\bibfnamefont{D.}~\bibnamefont{Van~Harlingen}},
  \bibinfo{journal}{Rev. Mod. Phys.} \textbf{\bibinfo{volume}{67}},
  \bibinfo{pages}{515} (\bibinfo{year}{1995}).

\bibitem[{\citenamefont{Graf et~al.}(1996)\citenamefont{Graf, Yip, Sauls, and
  Rainer}}]{graf1996}
\bibinfo{author}{\bibfnamefont{M.~J.} \bibnamefont{Graf}},
  \bibinfo{author}{\bibfnamefont{S.}~\bibnamefont{Yip}},
  \bibinfo{author}{\bibfnamefont{J.}~\bibnamefont{Sauls}}, \bibnamefont{and}
  \bibinfo{author}{\bibfnamefont{D.}~\bibnamefont{Rainer}},
  \bibinfo{journal}{Phys. Rev. B} \textbf{\bibinfo{volume}{53}},
  \bibinfo{pages}{15147} (\bibinfo{year}{1996}).

\bibitem[{\citenamefont{Matsuda et~al.}(2006)\citenamefont{Matsuda, Izawa, and
  Vekhter}}]{matsuda2006}
\bibinfo{author}{\bibfnamefont{Y.}~\bibnamefont{Matsuda}},
  \bibinfo{author}{\bibfnamefont{K.}~\bibnamefont{Izawa}}, \bibnamefont{and}
  \bibinfo{author}{\bibfnamefont{I.}~\bibnamefont{Vekhter}},
  \bibinfo{journal}{J. Phys.: Condensed Matter} \textbf{\bibinfo{volume}{18}},
  \bibinfo{pages}{R705} (\bibinfo{year}{2006}).

\end{thebibliography}
%

\end{document}